\documentclass[a4paper,11pt]{article}
\usepackage{authblk}
\usepackage{graphicx}
\usepackage{amsmath}
\usepackage{mathrsfs}
\usepackage{amssymb}
\begin{document}
\title{Probing short range correlations in Heavy-Ion Double Charge Exchange reactions
}

\author[1,2]{C. Garofalo}
\author[3]{H.Lenske}
\author[1,2]{F. Cappuzzello}
\author[2]{M. Cavallaro}
\affil[1]{Dipartimento di Fisica e Astronomia 'Ettore Majorana', Università di Catania, Catania, Italy}
\affil[2]{Istituto Nazionale di Fisica Nucleare, Laboratori Nazionali del Sud, Catania, Italy}
\affil[3]{Institut fur Theoretische Physik, Justus–Liebig–Universitat Giessen, D-35392 Giessen, Germany}


\date{}
\maketitle
\begin{abstract}
The Majorana Double Charge Exchange (MDCE) provides a suitable environment for studying the dynamics of the neutrinoless double beta ($0\nu\beta\beta$) decay, particularly short-range correlations among nucleons. The study of the pion potential is essential in this respect, as it represents the strong interaction counterpart of the neutrino potential, driving nucleon correlations in  $0\nu\beta\beta$ decay. Numerical studies on pion potential have revealed an effective range of about $1$ fm with slight dispersion around this value, confirming the short-range character of the MDCE process.

\end{abstract}

\section{Introduction}
NUMEN (NUclear Matrix Elements for Neutrinoless double beta decay) \cite{cappuzzello2023shedding, cappuzzello2018numen} is a project proposed in 2015 with the aim of studying Double Charge Exchange (DCE) reactions induced by heavy-ion, as they represent a special class of direct nuclear reactions suitable for spectroscopic research. In more detail, the DCE reactions are second-order nuclear reactions mainly fed through three main competitive processes, namely the Transfer Double Charge Exchange (TDCE) \cite{ferreira2022multinucleon, ferreira2025analysis}, the Double Single Charge Exchange (DSCE) \cite{bellone2025heavy, bellone2020two} and the Majorana Double Charge Exchange (MDCE) \cite{lenske2024theory}. The TDCE proceeds through the direct transfer of nucleons between the two nuclei, while DSCE and MDCE are processes in which two uncorrelated (the former) or correlated (the latter) nucleons interact by exchanging charged mesons. The MDCE represents a novel and promising approach for studying the dynamics of the neutrinoless double beta ($0\nu\beta\beta$) decay \cite{majorana1937teoria, furry1939transition}, since the same nuclear configurations can be explored using a transition operator that presents close mathematical similarities with the one driving the $0\nu\beta\beta$ decay \cite{lenske2024formal, vergados2016neutrinoless}. Indeed, both operators present Fermi, Gamow Teller, and rank-2 tensor components \cite{lenske2024theory, cappuzzello2018numen}.

\section{Majorana Double Charge Exchange mechanism}
\begin{figure}[htb]
\centerline{%
\includegraphics[width=5 cm]{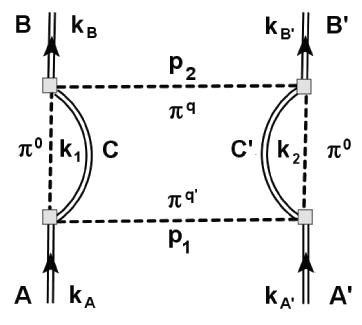}}
\caption{Scheme of the MDCE for the $A(A', B')B$ reaction \cite{lenske2024theory}.}
\label{Fig:MDCE}
\end{figure}

The box diagram of the MDCE process is presented in Fig. \ref{Fig:MDCE}. In the initial channel, the target $A$ and the projectile $A'$ interact by exchanging a charged pion $\pi^{\pm}$. The pion-nucleon scattering taking place in the two nuclei, leads to an excitation of $np^{-1}$ or $pn^{-1}$ Single Charge Exchange (SCE) particle–hole configurations \cite{lenske2023interactions} and the emission of a neutral pion $\pi^0$. Both intermediate nuclei $C, C'$ and $\pi^0$ propagate in the intermediate s-channel until a second charge exchange reactions by a $\pi^0 \rightarrow \pi^{\pm} $ process occurs, populating the exit channel nucleus $B(Z \pm 2)$ as discussed in detail in Ref. \cite{lenske2024theory}. Each of the SCE-type transition is driven by the pion-nucleon isovector T-matrix $T_{\pi N}$, which describes the pion-nucleon scattering taking place at each vertex of the box diagram.

\subsection{The pion potential}
Using the pion rest mass, $m_{\pi}\sim 139\,MeV$, as a natural separation scale \cite{lenske2024theory}, the pion potential $U_{\pi}$ is evaluated analytically in closed form, i.e., by replacing the sum over intermediate states with the identity operator, as commonly performed in the $0\nu\beta\beta$ framework \cite{sen2013neutrinoless, vsimkovic2011relation}. Within this approximation, the pion potential $U_{\pi}$ depends on the pion-nucleon isovector T-matrix $T_{\pi N}$ according to the following equation:

\begin{equation}
    U_{\pi}(x)= - \int \frac{d^3 k}{(2\pi)^3} T_{\pi N}(\vec{p}_2,\vec{k})\frac{e^{i\vec{k}\cdot\vec{x}}}{m_{\pi}^2+k^2} T_{\pi N}(\vec{p}_1,\vec{k}) \,\,\, ,
    \label{eq:U}
\end{equation}

where $x$ is the distance between the two nucleons participating in the DCE transition, $k$ is the invariant relative pion-nucleon momentum:

\begin{equation}
    k^2 = \frac{1}{4s_{\pi N}}   \left( s_{\pi N} - \left(m_N + m_{\pi} \right)^2 \right)\left( s_{\pi N} - \left(m_N - m_{\pi} \right)^2\right) \quad ,
    \label{eq:k^2}
\end{equation}

where $m_N$ is the mass of the nucleon while $s_{\pi N}$ is the energy per nucleon, defined as $s_{\pi N} = s_{\pi C} /A^2$, with $A$ the mass number of the intermediate nucleus $C$. \\
By expanding the $T_{\pi N}$ in partial waves, $T_{\pi N}$ can be written as a superposition of three terms describing the formation of a resonance of the S-wave ($T_0$) and P-wave ($T_1$ and $T_2$) type \cite{moorhouse1969pion, johnson1993pion} :

\begin{equation}
    T_{\pi N} = \left[T_0(s_{\pi N}) + \frac{1}{m_{\pi}^2}{\left( T_1(s_{\pi N})\vec{p}\cdot\vec{p}' + iT_2 (s_{\pi N}) \vec{\sigma} \cdot(\vec{p}\times p') \right)}\right] \vec{T}_{\pi} \cdot\vec{\tau_{N}} \,.
    \label{eq:T-matrix}
\end{equation}

In the expression (\ref{eq:T-matrix}), $p$ and $p'$ are the pion and nucleon momenta, respectively, $\vec{\sigma}$ is the spin operator while $\vec{T}_{\pi}$ and $\vec{\tau_{N}}$ are the pion and the nucleon isospin operators. \\
The pion potential $U_{\pi}$ consists of nine terms, as it arises from the product of two T-matrices $T_{\pi N}$, which themselves are superpositions of three terms (see Eq. (\ref{eq:T-matrix})). By choosing collinear exchange charged pion momenta, i.e. $\vec{p}_1 || \vec{p}_2$ and $|\vec{p}_1|= |\vec{p}_2|$, by symmetry the number of independent pion potential components reduces from nine to six independent ones. \\

\subsection{Pion potential as a strong counterpart of the $0\nu\beta\beta$ neutrino potential}

\begin{figure}[htb]
\centerline{%
\includegraphics[width=8 cm]{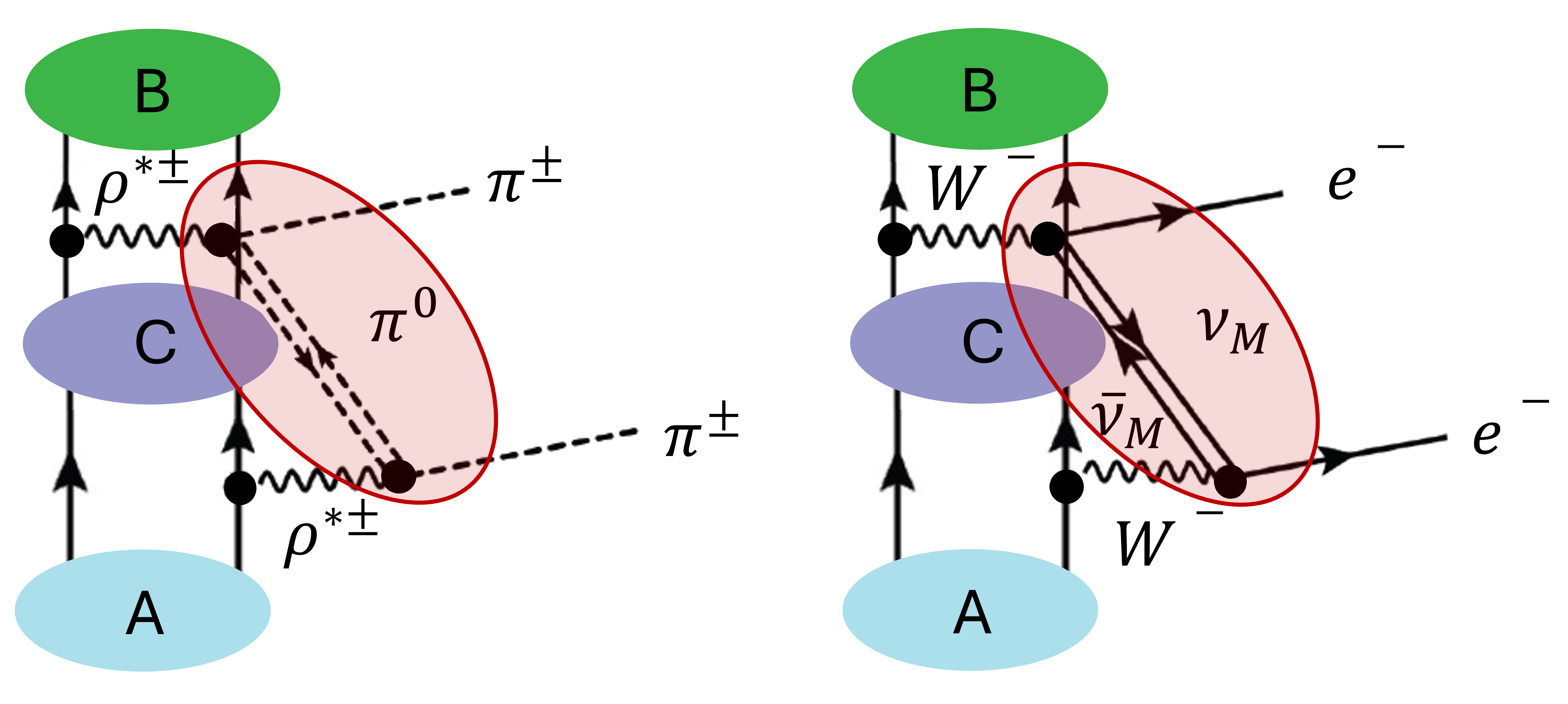}}
\caption{On the left, the hadronic MDCE for one of the interacting nuclei; on the right, the $0\nu\beta\beta$ process is illustrated.}
\label{Fig:MDCEandMDBD_conn}
\end{figure}

A direct comparison between the pion and neutrino potentials highlights several formal similarities, as both contain Fermi, Gamow–Teller, and tensor components \cite{cappuzzello2023shedding, iwata2017heavy}. 
In Fig. \ref{Fig:MDCEandMDBD_conn}, the formal similarities between the MDCE reaction (on the left) and the $0\nu\beta\beta$ decay (on the right) can be appreciated. As it can be seen, the same nuclei are involved and both processes proceed through an intermediate channel. Moreover, in each process, the two nucleons are correlated through either a neutral pion (in the MDCE) or a Majorana neutrino (in the $0\nu\beta\beta$ decay). In this respect, the pion potential represents the strong interaction counterpart of the Majorana neutrino potential. Thus, studying this object gives an insight into the $0\nu\beta\beta$ dynamics, highlighting the short-range correlations among nucleons. \\

\section{Results}
\begin{figure}[htb]
\centerline{%
\includegraphics[width=13.2 cm]{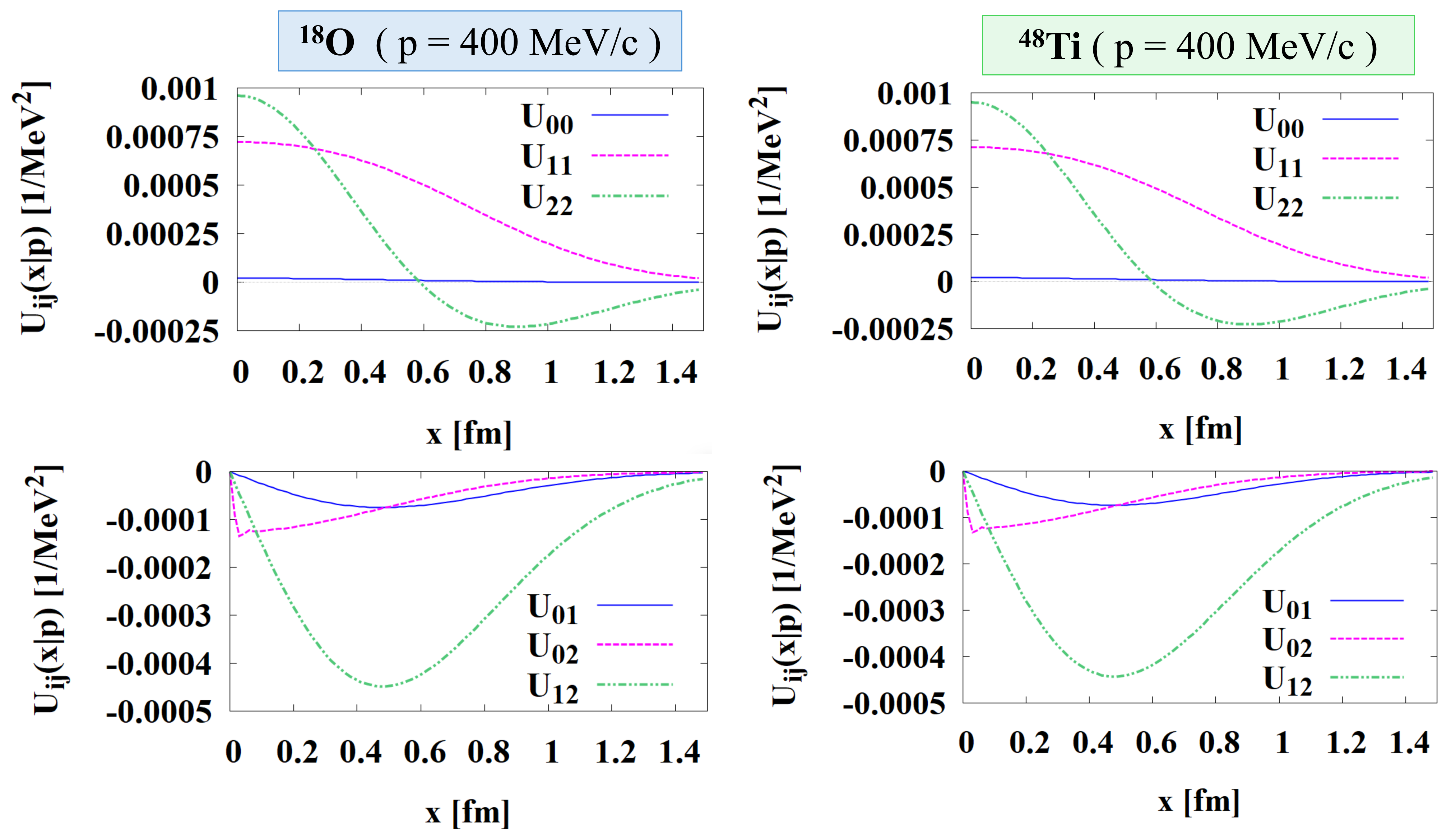}}
\caption{The diagonal (top panel) and non-diagonal (bottom panel) components of the pion potentials in the $^{18}$O and $^{48}$Ti collision at $T_{lab} = 270$ MeV are shown.}
\label{Fig:potential_18O48Ti}
\end{figure}

The pion potentials contributing to the reaction of a $^{18}$O beam colliding on the $^{48}$Ti target at $T_{lab} = 270$ MeV was investigated numerically. The pion potential as a function of the distance $x$ between the two nucleons involved in the MDCE reaction (exchanging a pion with momentum $p=400$ MeV/c) is shown in Fig. \ref{Fig:potential_18O48Ti} for both nuclei. The components $U_{ij}$ identify the S-, P-, and mixed contributions to the pion potential $U_\pi$,where the subscripts $i$ and $j$ label the form factors $T_i, T_j$ entering the expression in Eq. \ref{eq:U}. As it can been seen, the P-wave potential, i. e. the $U_{11}$ and $U_{22}$ components, dominate over the S-wave ones. Thus, it can be concluded that the MDCE process is dominated by the P-wave component.

Important global properties  of pion potentials like strength, range and spatial extension are
characterized by the monopole moments

\begin{equation}
    M^{(n)}_{ij}=\int d^3r r^n U_{ij}(\vec{r}) \quad,
\end{equation}

\noindent which were used to calculate the normalized radial moments
 
\begin{equation}
    \langle r^n\rangle_{ij} = \frac{M^{(n)}_{ij}}{M^{(0)}_{ij}} \quad.
\end{equation}

\noindent Of special interest are the $n=1,2$ moments allowing to determine the root-mean-square radii
$r^{(ms)}_{ij}=\sqrt{\langle r^2\rangle_{ij}}$ and the variances

\begin{equation}
    v_{ij}=\langle r^2\rangle_{ij}-\langle r\rangle^{2}_{ij} \quad.
\end{equation}

In Table \ref{tab:moments_norm}, the normalized linear radial monopole, the root-mean-square radii and the variance are reported for all components of the pion potential. Given the dominance of the P-wave component over the S-wave, the interaction is found to extend over a radius of approximately $1$ fm with a very small dispersion around the mean value. Indeed, the variances remain of moderate magnitude ranging from $0.03$ fm$^2$ for the P-wave component up to $0.10$ fm$^2$ for the S-wave component. These findings, indicating the very limited range of the pion potential, provide clear evidence of the markedly short-range character of the MDCE process. A direct consequence is the short-range nucleon–nucleon correlation, as the two nucleons participating in the DCE transition are connected through a neutral pion, whose potential exhibits a strongly short-range character. \\
The values obtained and reported in Table \ref{tab:moments_norm} comply perfectly well with the estimates obtained from the intranuclear momenta $\sim 100-200$ MeV/c ($r_m \sim 1-2$ fm) - used as the standard reference value in $0\nu\beta\beta$-papers \cite{tomoda1991double, vsimkovic20090} - and the average distances between nucleons in saturated nuclear matter ($r_0 \sim 1.14$ fm).

\begin{table}\begin{center}
\small
\caption{Table of the normalized linear radial monopole $\langle r \rangle_{ij}$, the root-mean-square radii $r^{(ms)}_{ij}$ and the variance $v$ for all the components of the pion potential.}
 \begin{center}
 \begin{tabular}{ c| c | c | c}
    \hline \hline
     Components &$\langle {r} \rangle_{ij}$ [fm] & $r^{(ms)}_{ij}$ [fm]   &$\nu_{ij}$ [fm$^2$]  \\ \hline 
     00 &$1.89$  & $2.53$  & $0.10$  \\ 
     11 &$1.04$  & $1.29$  & $0.03$  \\
     22 &$1.49$  & $1.70$  & $0.06$  \\
     01 &$1.41$  & $1.88$  & $0.05$  \\
     02 &$1.20$  & $1.64$  & $0.04$  \\
     12 &$1.06$  & $1.28$  & $0.03$  \\ \hline\hline
\end{tabular}
  \end{center}
 \label{tab:moments_norm}
 \end{center}
\end{table}

\section{Conclusions}
The MDCE represents a suitable tool for investigating the $0\nu\beta\beta$ dynamics, providing insight into short-range correlations and nucleon resonances, both of which are relevant for $0\nu\beta\beta$ decay. In this context, the study of the pion potential is of central interest, as it provides access to the short-range nucleon–nucleon correlations dominating the heavy-ion DCE reactions. In the present manuscript, numerical results for the range of the pion potential have been presented, proving the short-range character of the MDCE process. Indeed an effective range of approximately $1$ fm and a slight dispersion around this value was found. These results are of significant interest, since the pion potential can be regarded as the counterpart in the strong interaction sector of the neutrino potential, which is responsible for the correlations between the two nucleons exchanging the Majorana neutrinos in the $0\nu\beta\beta$ decay. Once experimental and theoretical studies become available, the pion potentials can be extracted from the cross sections. This will enable an independent investigation of nuclear isovector spectroscopy and a direct probe of short-range two-body correlations in nuclei, both of which are crucial for the study of $0\nu\beta\beta$ decay.\\

H. Lenske acknowledges financial support in part by INFN and DFG, grant Le439/7.



\bibliography{biblio}
\end{document}